%% file: main.tex
\documentclass[%
	reprint,
	superscriptaddress,
	amsmath,amssymb,amsfonts,
	aps,
	pra,
]{revtex4-2}

\usepackage[utf8]{inputenc}
\usepackage[T1]{fontenc}
\usepackage{lmodern}
\usepackage[english]{babel}
\usepackage[autostyle=true]{csquotes}

\usepackage[dvipsnames]{xcolor}
\usepackage{graphicx}
\graphicspath{{figures/}}

\usepackage{tikz}
\usepackage{tikz-cd}
\usepackage[normalem]{ulem}
\usepackage{mhchem}
\usetikzlibrary{%
  matrix,%
  calc,%
  arrows%
}
\usetikzlibrary{quantikz2} 

\usepackage{amssymb,amsmath}

\usepackage{bm}
\usepackage{mathtools}
\usepackage{braket}

\usepackage{xspace}
\usepackage{extdash}
\usepackage{mleftright}
\mleftright

\usepackage{hyperref}
\hypersetup{
}

\usepackage[
	capitalise,
]{cleveref}

\usepackage{acro}
\acsetup{%
	first-style=long-short,
}

\usepackage{url}

\definecolor{mypurple}{rgb}{0.49,0.18,0.56}
\definecolor{mygold}{rgb}{0.93,0.49,0.13}
\definecolor{mygreen}{rgb}{0,0.5,0}
\definecolor{myblue}{rgb}{0,0,0.75}
\definecolor{mymagenta}{cmyk}{0,1,0,0.12}
\definecolor{mygray}{rgb}{0.5,0.5,0.5}

\newif\ifcomments
\commentstrue 

\newcommand*{\bigO}{\mathcal{O}}

\renewcommand{\vec}{\bm}

\begin{document}
\title{Enhancing Expressivity of Quantum Neural Networks Based on the SWAP test}

\author{Sebastian Nagies}
\email{sebastian.nagies@unitn.it}
\thanks{Corresponding author}
\affiliation{Pitaevskii BEC Center and Department of Physics, University of Trento, Via Sommarive 14,
38123 Trento, Italy}
\affiliation{INFN-TIFPA, Trento Institute for Fundamental Physics and Applications, Trento, Italy}

\author{Emiliano Tolotti}
\affiliation{Department of Information Engineering and Computer Science, University of Trento,
Via Sommarive 9, 38123 Trento, Italy}

\author{Davide Pastorello}
\affiliation{Department of Mathematics, Alma Mater Studiorum - University of Bologna
piazza di Porta San Donato 5, 40126 Bologna, Italy}
\affiliation{INFN-TIFPA, Trento Institute for Fundamental Physics and Applications, Trento, Italy}

\author{Enrico Blanzieri}
\affiliation{Department of Information Engineering and Computer Science, University of Trento,
Via Sommarive 9, 38123 Trento, Italy}
\affiliation{INFN-TIFPA, Trento Institute for Fundamental Physics and Applications, Trento, Italy}

\begin{abstract}
Quantum neural networks (QNNs) based on parametrized quantum circuits are promising candidates for machine learning applications, yet many architectures lack clear connections to classical models, potentially limiting their ability to leverage established classical neural network techniques. We examine QNNs built from SWAP test circuits and discuss their equivalence to classical two-layer feedforward networks with quadratic activations under amplitude encoding. Evaluation on real-world and synthetic datasets shows that while this architecture learns many practical binary classification tasks, it has fundamental expressivity limitations: polynomial activation functions do not satisfy the universal approximation theorem, and we show analytically that the architecture cannot learn the parity check function beyond two dimensions, regardless of network size. To address this, we introduce generalized SWAP test circuits with multiple Fredkin gates sharing an ancilla, implementing product layers with polynomial activations of arbitrary even degree. This modification enables successful learning of parity check functions in arbitrary dimensions as well as binary n-spiral tasks, and we provide numerical evidence that the expressivity enhancement extends to alternative encoding schemes such as angle (Z) and ZZ feature maps. We validate the practical feasibility of our proposed architecture by implementing a classically pretrained instance on the IBM Torino quantum processor, achieving 84\% classification accuracy on the three-dimensional parity check despite hardware noise. Our work establishes a framework for analyzing and enhancing QNN expressivity through correspondence with classical architectures, and demonstrates that SWAP test-based QNNs possess broad representational capacity relevant to both classical and potentially quantum learning tasks.
\end{abstract}

\maketitle

\section{Introduction}\label{sec:intro}
\input{Introduction}

\section{Constructing quantum neural networks with SWAP tests}\label{sec:background}
\input{background}

\section{Assessing expressivity on classical datasets}\label{sec:numerics}

\input{numerics}

\section{Implementation on quantum hardware}\label{sec:implementation}

\input{implementation}

\section{Conclusions}\label{sec:conclusion}

\input{Conclusion}

\section*{Code Availability}
Code and data used in this work are available: \url{https://github.com/snagies/generalized-swap-test-qnn}.

\input{Acknowledgments}

\appendix

\input{appendix}

\bibliographystyle{apsrev4-2}
\bibliography{qml}

\end{document}

%% file: Introduction.tex
Quantum machine learning (QML) is a rapidly growing field that develops quantum algorithms for machine learning tasks \cite{2014,lloyd_quantum_2013,biamonte_quantum_2017,Schuld2021,Pastorello2023}. Quantum mechanics provides unique computational resources, including entanglement and non-stabilizerness, that may enable advantages over classical computation. These potential advantages have been demonstrated in several contexts, such as the quantum support vector machine \cite{Rebentrost2014}.

Quantum neural networks (QNNs) represent a key component of QML, aiming to extend the remarkable success of classical neural networks to the quantum domain. QNNs are hybrid quantum-classical systems in which the weights of a parametrized quantum circuit are adjusted through classical optimization of an objective function, based on measurement outcomes obtained from the circuit \cite{schuld_circuitcentric_2020,cerezo_variational_2021}.
Two main architectural approaches have emerged. The first and most widely adopted approach employs variational quantum circuits (VQCs) \cite{peruzzo_variational_2014}, using parametrized circuits tailored to specific problems or hardware constraints \cite{Farhi2014,Cincio2018,Rattew2020}, with the Hardware Efficient Ansatz \cite{Kandala2017} being a prominent example. 
The second approach focuses on designing architectures that closely resemble classical neural networks by generalizing fundamental building blocks, such as the perceptron, to the quantum circuit setting \cite{Tacchino2019,tacchino_quantum_2020}. This latter approach holds the potential of systematically translating established classical neural network techniques and results to the quantum realm. In both cases, fundamental limitations such as the barren plateau problem \cite{mcclean_barren_2018,Cerezo2023,ragone_lie_2024} must be addressed before these systems can become practically useful.

This paper focuses on QNN architectures constructed from SWAP test quantum circuits \cite{zhao_building_2019,li_quantum_2020,pastorello_scalable_2024} and addresses a specific limitation: their expressivity on classical datasets, which may have implications for their performance on quantum learning tasks as well.

The SWAP test \cite{buhrman_quantum_2001} measures the overlap between two arbitrary quantum states and, when processing classical data via amplitude encoding \cite{schuld_supervised_2018}, is functionally equivalent to a classical perceptron with a quadratic activation function. This correspondence enables the construction of quantum perceptrons as modular building blocks for QNNs. Quadratic activation functions have received renewed interest in classical machine learning due to their favorable optimization landscapes and approximation properties \cite{livni_computational_2014, Fan2018,mannelli_optimization_2020}. Moreover, such activation functions arise naturally in other quantum machine learning contexts: they appear in quantum kernel methods \cite{Schuld2021a}, and recently in quantum optical classifiers based on Hong-Ou-Mandel interference \cite{Roncallo2025}.

The SWAP test can be implemented across diverse quantum computing platforms: it can be decomposed into platform-specific native gate sets for superconducting qubits \cite{Cincio2018,Arute2019, Bravyi2024}, trapped ions \cite{Piltz2016,Chen2024, Liu2025, Meth2025}, and neutral atoms \cite{Brodoloni2025, GonzalezCuadra2025}. Alternatively, the SWAP test can be directly implemented in experiment, as demonstrated with optical platforms \cite{Fiorentino2005, Kang2019, Baldazzi2024} and trapped ions \cite{Linke2018, Nguyen2021}. This broad compatibility makes the SWAP test a suitable building block for near-term quantum neural networks.

In Ref.~\cite{pastorello_scalable_2024} a two-layer feedforward neural network constructed exclusively from SWAP test quantum circuits was proposed. These modules execute SWAP tests between amplitude-encoded input and weight vectors, realizing quadratic activation functions while potentially requiring only a small number of qubits. However, feedforward neural networks with a single hidden layer and quadratic activation functions do not satisfy the universal approximation theorem \cite{Hornik1989, hornik_approximation_1991}, which implies that certain functions cannot be approximated regardless of network width. Deep architectures, and thus large circuit depths, would be required to overcome this limitation, which is not compatible with current noisy quantum hardware.

This work generalizes the modular SWAP test-based QNN architecture to enable substantially more expressive quantum neural networks, without the need for significantly larger circuit depth. We propose using generalized SWAP test circuits as building blocks: these circuits employ multiple Fredkin gates controlled by a single shared ancilla qubit, acting on multiple pairs of inputs and weights. This construction effectively increases the degree of the polynomial activation function, enabling architectures that closely resemble classical polynomial neural networks \cite{livni_computational_2014,blondel_polynomial_2016} or neural networks with product layers \cite{shin_pisigma_1991,li_sigmapisigma_2003}. The degree of the activation function can be scaled arbitrarily, enhancing network expressivity as demonstrated through our numerical experiments. In particular, our architecture can approximate complex functions, such as the parity check in arbitrary dimensions, that are provably impossible to learn with the original design beyond two dimensions, regardless of network size. We provide an analytical argument establishing this fundamental limitation of the quadratic architecture and demonstrate that our generalization overcomes it. We complement our theoretical and numerical analysis with an experimental demonstration on IBM's Torino quantum processor (Heron r1). A classically pretrained instance of our architecture successfully classifies the three-dimensional parity check with 84\% accuracy, indicating that the approach tolerates the inherent gate errors and decoherence of current quantum hardware. Finally, we investigate the generality of our QNN by extending the architecture beyond amplitude encoding to alternative data encoding schemes, such as phase encoding. While the mathematical correspondence to classical feedforward neural networks is specific to amplitude encoding, our numerical results provide evidence that the generalized SWAP test circuit enhances expressivity also under these alternative encodings for certain problem classes.

This paper is organized as follows: Section \ref{sec:background} explains how QNNs can be constructed from SWAP tests and introduces a generalized quantum circuit architecture that effectively implements a product layer. Section \ref{sec:numerics} presents numerical results demonstrating our architecture's performance on both classical real-world datasets and the challenging synthetic parity check function and binary spiral task. Section \ref{sec:implementation} validates the practical feasibility of our approach by running a pretrained model on real quantum hardware to learn the parity check function. Section \ref{sec:conclusion} summarizes our findings and conclusions.

%% file: background.tex
The quantum SWAP test is a fundamental operation in quantum computing which allows for the estimation of the overlap between two arbitrary quantum states. When representing classical vectors with those quantum states via amplitude encoding, the output of the SWAP test is functionally equivalent to the output of a classical perceptron used in feedforward neural networks. Whereas many parametrized quantum circuits have little resemblance to classical neural networks, a quantum neural network composed of SWAP test circuits as its building blocks has a one-to-one correspondence with its classical counterparts \cite{pastorello_scalable_2024}. Given the extraordinary success of classical neural networks for machine learning tasks, this QNN architecture potentially holds great promise for quantum learning applications as well. 

This section provides a brief review of the SWAP test quantum circuit and demonstrates how quantum feedforward neural networks can be constructed from these building blocks. We discuss the universality of this architecture and introduce a simple generalization in Section \ref{subsec:productlayer} that significantly enhances expressivity on certain classical datasets (see Section \ref{sec:numerics}). At the end, we comment on the architecture's scalability on current quantum hardware in Section \ref{subsec:scalibility}.

\subsection{SWAP test}\label{subsec:swap_test}

\begin{figure}[h]
    \centering
    \resizebox{0.8\columnwidth}{!}{
    \begin{quantikz}[column sep=9pt, row sep={18pt,between origins}]
        \lstick{$\ket{0}$} & & \gate{H} & \ctrl{2} &  \gate{H} & \meter{} & \setwiretype{c} \\
        \lstick{$\ket{\psi}$} & \qwbundle{\delta} & & \targX{} & & & \\
        \lstick{$\ket{\phi}$} & \qwbundle{\delta} & & \targX{} & & &
    \end{quantikz}
    }
    \caption{Quantum circuit implementing the SWAP test which estimates the overlap between two quantum states $\ket{\psi}$ and $\ket{\phi}$ by measuring an ancilla qubit initialized to state $\ket{0}$. Both states are represented on quantum registers with $\delta$ qubits. If the two quantum states encode classical inputs and weights via amplitude encoding in a quantum perceptron context, the number of required qubits is $\delta = \lceil\log_2{d}\rceil$, where $d$ is the dimension of the classical input and weights vectors.}
    \label{fig:swap_module}
\end{figure}
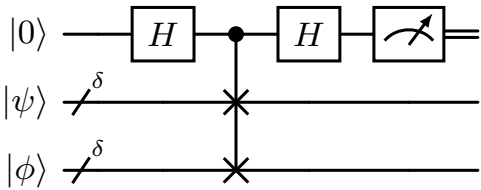

 Given two quantum states $|\psi\rangle$ and $|\phi\rangle$, the SWAP test circuit operates as follows (see Fig. \ref{fig:swap_module}):

\begin{enumerate}
    \item Initialize an ancilla qubit in the $|0\rangle$ state.
    \item Apply a Hadamard gate to the ancilla qubit.
    \item Apply a controlled-SWAP operation, with the ancilla qubit as the control and $|\psi\rangle$ and $|\phi\rangle$ as the targets.
    \item Apply another Hadamard gate to the ancilla qubit.
    \item Measure the ancilla qubit in the computational basis.
\end{enumerate}

 The probability $P$ of measuring the ancilla qubit in the $|0\rangle$ state is then given by:

\begin{equation}\label{eq:swap_test}
    P(0) = \frac{1}{2}(1 + |\langle\psi|\phi\rangle|^2).
\end{equation}

This probability is directly related to the overlap between the two input states: For two orthogonal (identical) states the probability is $P(0) = 0.5$ ($  P(0) = 1$). Through repeated preparation of the input states and measurement of the ancilla qubit after the SWAP test, this probability (and thus the overlap between the two states) can be estimated to arbitrary precision $\epsilon$ with $\bigO (\epsilon^{-2})$ samples.

On certain quantum computing platforms, such as photonic or trapped ion quantum computers, the SWAP test can be implemented natively \cite{Fiorentino2005, Kang2019, Baldazzi2024,Linke2018, Nguyen2021}, thus making it an attractive building block for quantum machine learning applications. On other platforms (e.g. superconducting qubits) the SWAP test needs to be decomposed into the respective native gate set first.

\subsection{Quantum perceptron}\label{subsec:perceptron}

A classical perceptron typically takes as input a classical $d$-dimensional vector $\vec{x}$, is parametrized by a $d$-dimensional weight vector $\vec{w}$ and a bias $b$, and outputs a single number $m(\vec{x}; \vec{w}, b)$. The whole procedure can be written as

\begin{align}\label{eq:classical_preceptron}
    m(\vec{x}; \vec{w}, b) = \sigma( \vec{x} \cdot \vec{w} + b),
\end{align}
where $\cdot$ denotes the dot product and $\sigma$ is a nonlinear activation function. Common choices for $\sigma$ include non-polynomial functions like the sigmoid or ReLU. However, polynomial activation functions, particularly quadratic ones ($\sigma(z) = z^2$), have also been investigated \cite{livni_computational_2014,mannelli_optimization_2020} and are relevant to the quantum setting discussed in this work.

Comparing with Eq. \ref{eq:swap_test}, we observe that the output of the SWAP test is formally similar to that of a classical perceptron. Specifically, the overlap $|\langle\psi|\phi\rangle|^2$ can be identified with a classical cosine similarity pre-activation paired with a quadratic activation function. 

Cosine similarity is sometimes used in classical perceptrons as an alternative to the simple dot product, as it provides a bounded pre-activation value between -1 and 1 and achieves better performance in certain contexts \cite{Luo2017}:

\begin{align}
    \cos(\vec{x}, \vec{w}) = \frac{\vec{x} \cdot \vec{w}}{||\vec{x}|| \cdot||\vec{w}||}.  
\end{align}

To establish the link to the inner product of quantum states $\langle\psi|\phi\rangle$, we first encode a $d$-dimensional classical input $\vec{x}$ into the quantum state $\ket{\psi}$ via amplitude encoding \cite{schuld_supervised_2018}:

\begin{equation}\label{eq:amplitude_encoding}
    |\psi\rangle = \frac{1}{\|\vec{x}\|} \sum_{i=1}^{2^\delta} x_i |i\rangle,
\end{equation}

where $\delta = \lceil\log_2{d}\rceil$ is the number of qubits needed for the encoding and $\ket{i}$ denotes the $i$th computational basis state. After encoding the weights $\vec{w}$ analogously into $\ket{\phi}$, the inner product of the quantum states is equivalent to the classical cosine similarity.

To make the correspondence to the classical case complete, one would also need a bias parameter which offsets the result of the inner product before passing it to the activation function (squaring it in this case). Although we can not implement this directly with the here considered architecture, we can achieve something similar by introducing a dummy input feature: Instead of a $d$-dimensional input $\vec{x}$, we use a $(d+1)$-dimensional input $\vec{x}^\prime$, where we set the last feature to 1. The last parameter $w_{d+1}$ of the $(d+1)$-dimensional weight vector $\vec{w}^\prime$ than acts as an effective bias and the output of the SWAP test (again assuming amplitude encoding) is given by 

\begin{align} \label{eq:quantum_bias}
    P(0) = \frac{1}{2}\left(1 + \left|\frac{\vec{x} \cdot \vec{w} + w_{d+1}}{||\vec{x}^\prime|| \cdot||\vec{w}^\prime||} \right|^2\right).
\end{align}

This construction is not completely equivalent to the classical perceptron (Eq. \ref{eq:classical_preceptron}), as the bias in this case is included in $\vec{w}^\prime$ and is thus part of the pre-activation function ($||\vec{x}^\prime|| = \sqrt{||\vec{x}||^2 + 1}$ and $||\vec{w}^\prime|| = \sqrt{||\vec{w}||^2 + w_{d+1}^2}$). Nevertheless, as our numerical results in Sec. \ref{sec:numerics} show, this architecture is still suitable to learn classical datasets.

Finally, we note that although amplitude encoding provides exponential compression in qubit count, generic state preparation requires exponentially many gates \cite{Plesch2011}, potentially negating quantum advantages. Quantum random access memory (QRAM) architectures, most notably the bucket-brigade design, can in principle achieve state preparation with $\bigO(\log N)$ query complexity, but remain experimentally challenging \cite{Giovannetti2008,Xu2023, Berti2025, Shen2025}. For near-term devices, approximate methods have proven practical: variational approaches optimize parametrized circuits to approximate target states \cite{Nakaji2022}, while matrix product state techniques exploit limited entanglement to achieve efficient preparation with bounded error \cite{Malz2024}. For the SWAP test-based architecture discussed in this work, high-dimensional inputs can be partitioned onto multiple modules as discussed in Sec.~\ref{subsec:scalibility} and Appendix \ref{app:mnist}, reducing the state preparation complexity on each individual module.

\subsection{Two-layer feedforward neural network}\label{subsec:twolayer_nn}

Similarly to classical neural networks, quantum perceptrons based on the SWAP test can be combined to form a feedforward neural network. We consider a two-layer quantum-classical hybrid architecture for binary classification tasks, where the network's output is computed as:

\begin{align}\label{eq:standard_qnn}
    f(\vec{x}; \{\vec{w}\}) &= \sum_{i=1}^N c_i P_i(0) + b \nonumber \\
    &= \sum_{i=1}^N  \frac{c_i}{2}\left(1 + |\langle\vec{x}^\prime|\vec{w}_i^\prime\rangle|^2\right) + b.
\end{align}
Here, the classical $d$-dimensional input vector $\vec{x}$ is encoded into a quantum state $\ket{\vec{x}^\prime}$ via amplitude encoding (Eq.~\ref{eq:amplitude_encoding}). The prime notation denotes that the input contains a dummy feature to realize a bias term (see previous section). The state $\ket{\vec{x}^\prime}$ serves as the input to $N$ distinct SWAP tests in the first layer. Each SWAP test $i$ utilizes a unique classical weight vector $\vec{w}_i$ as well as a bias, both of which are encoded together into the quantum state $\ket{\vec{w}_i^\prime}$. In the rest of the article we will drop the prime notation and always assume that the input contains an additional dummy feature and one of the weights acts as a bias. 

The $N$ SWAP tests can be executed in parallel. Alternatively, they can be performed sequentially on the same quantum circuit, requiring repeated initialization of the weight and input states for each SWAP test. The first layer's outputs are the $N$ probabilities, $P_i(0)$, obtained from measuring the ancilla qubit of each SWAP test in the state $\ket{0}$.

In the second, purely classical layer, these $N$ probabilities are multiplied by the respective coefficient $c_i$ and then summed to produce a single scalar output. Finally we also add a classical bias $b$ in the second layer, which can be convenient for shifting the output of the network to fit with a chosen loss function. Overall, the network has $N(d+2) + 1$ trainable parameters, consisting of $Nd$ weights (from the $N$ $d$-dimensional vectors $\vec{w}_i$), $N$ biases in the quantum layer, $N$ classical coefficients $c_i$ and a single classical bias in the second layer.

Crucially, this network architecture does not satisfy the universal approximation theorem (UAT). The standard UAT typically requires non-polynomial activation functions~\cite{hornik_approximation_1991,pinkus_approximation_1999}, whereas the activation function implicit in $|\langle\vec{x}|\vec{w}_i\rangle|^2$ is polynomial (quadratic) in the components of $\vec{x}$ and $\vec{w}_i$. Furthermore, as discussed in the last section, the way we implement the bias is not equivalent to a bias term in a classical perceptron. In Sec.~\ref{sec:numerics}, we will empirically demonstrate that this architecture is nevertheless suitable for learning many real-world datasets. However, as shown with the parity check example in Sec.~\ref{sec:numerics}, certain challenging classical functions are impossible for this neural network to learn. Successfully learning such functions necessitates modifications to the quantum architecture. Several proposals exist in the literature for realizing non-polynomial activation functions (e.g., sigmoid) on quantum hardware~\cite{Tacchino2019, tacchino_quantum_2020, Huber2021}. However, in the next section we will introduce a simple modification to the current quantum neural network which allows us to stick with the SWAP test based architecture and is designed to enable the learning of these more challenging classical functions.

\subsection{Constructing a product layer}\label{subsec:productlayer}

The major drawback of the two-layer feedforward quantum neural network architecture discussed in the previous section is the quadratic activation function which limits the networks capability to learn certain classical datasets like the parity check. However, we can slightly modify the original SWAP test quantum circuit in order to increase the degree of the polynomial activation function to arbitrary even degrees.

\begin{figure}[h]
    \centering
    \begin{quantikz}[column sep=9pt, row sep={18pt,between origins}]
        \lstick{$\ket{0}$} & & \gate{H} & \ctrl{2} & \ \ldots \ & \ctrl{5} & \gate{H} & \meter{} & \setwiretype{c} \\
        \lstick{$\ket{\vec{x}}$} & \qwbundle{\delta} & & \targX{} & & & & & \rstick[5]{$\times k$}\\
        \lstick{$\ket{\vec{w}_{i1}}$} & \qwbundle{\delta} & & \targX{} & & & & &\\
        \setwiretype{n} \ldots & & & & \ldots\\
        \lstick{$\ket{\vec{x}}$} & \qwbundle{\delta} & & & & \targX{} & & &\\
        \lstick{$\ket{\vec{w}_{ik}}$} & \qwbundle{\delta} & & & & \targX{} & & &
    \end{quantikz}
    \caption{Generalization of the SWAP test to a product module. $w_{ij}$ is the weight vector in product module $i$, with factor index $j$. $\delta = \lceil\log_2{(d+1)}\rceil$ is the number of qubits needed 
    to amplitude encode the input and weight vectors $\vec{x}$ and  $\vec{w}_{ij}$ of dimension $d$ and $d+1$ respectively (assuming one of the weights acts as a bias). The overall product module can be seen as a number of $k$ SWAP tests (see Fig. \ref{fig:swap_module}) being executed using the same ancilla qubit. After measuring said ancilla, the probability $P(0)_i$ corresponds to a polynomial activation function of degree $2k$ (see Eq. \ref{eq:product_layer}).}
    \label{fig:product_module}
\end{figure}
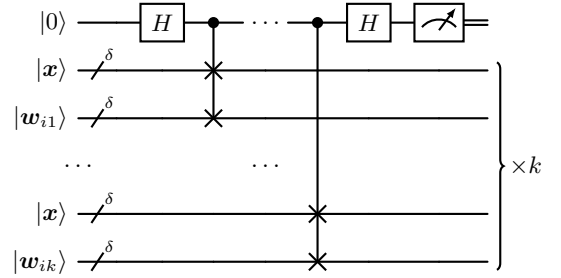

Our proposed generalization of the SWAP test is depicted in Fig. \ref{fig:product_module}. Whereas in the original SWAP test (see Fig. \ref{fig:swap_module}) the circuit was composed of one copy of input state $\ket{\vec{x}}$ and weights $\ket{\vec{w}_i}$ as well as a single ancilla qubit, here we consider a number of $k$ copies of the input state as well as $k$ (generally different) states $\ket{\vec{w}_{ij}}$. The difference now lies in the fact that all $k$ pairs of inputs and weights (which we refer to as \emph{factor modules}) share the same single ancilla qubit (again initialized to $\ket{0}$), i.e. we perform $k$ SWAP tests using the same ancilla qubit before measuring its probability $P(0)_i$. Analogous to the standard setup (Eq. \ref{eq:standard_qnn}), we have $N$ different instances of these \emph{product modules}.  

In the second layer the output probabilities get again multiplied by classical coefficients $c_i$ and summed up. The overall output of this generalized two-layer neural network (the first layer is a quantum version of a \emph{product layer}, requiring $N$ measurements, while the second layer is purely classical), can be computed as

\begin{align}\label{eq:product_layer}
    f(\vec{x}; \{\vec{w}\}) &= \sum_{i=1}^N c_i P(0)_i + b\nonumber\\
    &= \sum_{i=1}^N \frac{c_i}{2} \left(1 + \prod_{j=1}^k |\langle\vec{x}|\vec{w}_{ij}\rangle|^2 \right) + b.
\end{align}

Here we have a total of $N$ classical coefficients $c_i$, a single classical bias $b$, $Nk$ biases and $Nkd$ weights in the quantum layer, for an overall number of $N[k(d+1) + 1] + 1$ trainable parameters.

For $k=1$ this modified neural network architecture recovers the original two-layer feedforward network discussed in Sec. \ref{subsec:twolayer_nn}. Note that for $k > 1$, we have two possibilities: First, all the weight vectors $\ket{\vec{w}_{ij}}$ for a given product module with index $i$ can be chosen equal. In this case the network is equal to the standard two-layer feedforward network architecture in Eq. \ref{eq:standard_qnn}, but with an activation function of degree $2k$ instead of quadratic. However, we can also allow the weight vectors to be different from each other within the same product module, in which case the network has a similar structure to so called sigma-pi-sigma networks \cite{shin_pisigma_1991,li_sigmapisigma_2003}.

Formally, the above proposed quantum neural network architecture still does not satisfy the universal approximation theorem, as the activation function remains polynomial. However, as we can increase the (even) degree of the activation function arbitrarily by increasing $k$, we conjecture that this will become irrelevant for large networks and real-world datasets. Furthermore, for more general classical neural networks of sigma-pi-sigma type, which are structurally similar to our QNN architecture, universal approximation results, which don't require a non-polynomial activation function, are discussed in Refs.~\cite{luo_sup_2000,long_uniform_2007}. 

We note that the original SWAP test-based QNN architecture (Eq.~\eqref{eq:standard_qnn}), as well as its generalization to a QNN with product layer (Eq.~\ref{eq:product_layer}), can be thought of as parametrized quantum circuits with \emph{parallel encodings}, where the data encoding circuit is repeatedly applied over multiple subsystems. It was shown that such redundancy can increase a quantum circuits expressivity in Refs.~\cite{GilVidal2020,Schuld2021b}.

In Sec. \ref{sec:numerics} we give numerical evidence showing that, by increasing the number of product modules as well as the number $k$ of factor modules contained within each of them, the proposed architecture can indeed learn certain hard classical datasets, like the parity check and the $n$-spiral task, in arbitrary dimensions.

\subsection{Scalability of the quantum neural network }\label{subsec:scalibility}

A key bottleneck when implementing the proposed quantum neural network architecture on quantum hardware is the number of available qubits for currently realizable quantum SWAP tests. Some classical datasets (and similar considerations apply to quantum data) have a large dimension $d$, e.g., image classification problems. In those cases, potentially there aren't enough available qubits to encode the whole vector into a single quantum state. In Ref.~\cite{pastorello_scalable_2024} the authors explain how in such cases the input can be split onto multiple SWAP test modules. For the original quantum neural network (without product layers, see Sec. \ref{subsec:twolayer_nn}) the output of the network is then modified to 

\begin{align}\label{eq:scalable_qnn}
    f(\vec{x}; \{\vec{w}\}) = \sum_{i=1}^N  \frac{c_i}{2}\left(1 + |\langle\vec{x}^{(i)}|\vec{w}_i\rangle|^2\right) + b,
\end{align}

where $\vec{x}^{(i)}$ amplitude encodes only a subset of the features in the original input vector $\vec{x}$. The subset of features each different SWAP test receives can be varied for each module. Furthermore the subsets are also allowed to overlap or repeat across modules. This strategy can of course be equally applied to the modified network with product layers discussed in Sec. \ref{subsec:productlayer}.
In this case, the input features can also be split across the $k$ factor modules within each product module, instead of giving each factor module the same input, which allows for further generalization of the QNN architecture.

In Appendix \ref{app:mnist} we demonstrate with the example of the MNIST dataset \cite{lecun_gradientbased_1998} that the quantum neural network architecture defined in Eq. \ref{eq:product_layer} can reliably learn higher-dimensional input data, even when the features are split into multiple generalized SWAP tests.

%% file: numerics.tex
In this section we demonstrate the capability of our proposed quantum neural network architecture, based on the generalized SWAP test circuit (see Sec. \ref{subsec:productlayer}) to learn classical datasets. This expressivity for classical learning tasks is an important prerequisite for translating the impressive applications and successes of classical neural networks to quantum learning tasks.

We start in Sec.~\ref{subsec:training} with explaining our numerical implementation of the architecture and the metrics we use for quantifying its expressivity. We train the network to accurately classify 21 different real-world data sets in Sec.~\ref{subsec:real_world}. In Appendix \ref{app:mnist} we extend our numerical analysis to the high-dimensional MNIST dataset of handwritten digits, where features can be split onto multiple modules, further demonstrating the advantage of the modular approach of our proposed QNN architecture.
We consider the harder-to-learn parity check function in Sec.~\ref{subsec:parity_check}. Here we give an analytical argument for why the original two-layer network (without product layer) can not learn the higher-dimensional parity check and then present our numerical results for the generalized architecture, which strongly suggest that our proposed modified architecture can learn the parity check function in arbitrary dimensions. In Sec.~\ref{subsec:spiral} we show analogous numerical results for successfully learning another difficult synthetic dataset: the binary n-spiral task, demonstrating that our proposed architecture is agnostic to the specific problem structure. Finally, in Sec.~\ref{subsec:encodings} we discuss a further generalization of our architecture by considering other encoding schemes beyond amplitude encoding.

\subsection{Implementation and training} \label{subsec:training}

In an experimental implementation of the quantum neural network architecture proposed in the last section, the first layer of the network would be run on a quantum computer (initializing input, executing SWAP tests, measurement of ancilla qubits), while the second layer is computed classically. For most of this work we simulate the entire process purely classically (which is feasible for the considered classical datasets and network sizes), i.e., the output probabilities after the first quantum layer (either the standard version or with product layers, see Eqs. \ref{eq:standard_qnn} and \ref{eq:product_layer}) are computed exactly instead of sampling them from repeated real measurements. The only exception to this occurs in Sec. \ref{sec:implementation}, where we run a pretrained product layer quantum neural network on real quantum hardware.

For training the neural networks, 
we implemented a classical surrogate of the QNN in Python with the PyTorch library,
which allows us to effectively implement and extensively test the proposed architecture on a GPU.
The surrogate outputs the same probabilities as the quantum circuit.
Specifically, we implement the single hidden-layer QNN defined in Eq. \ref{eq:product_layer},
where we rescale
the output probabilities to be in the range $[0,1]$ for convenience, i.e.
the PyTorch QNN output is defined as
\begin{align}\label{eq:product_layer_pytorch}
    f(\vec{x}; \{\vec{w}\}) &= \sum_{i=1}^N c_i (2P(0)_i-1) + b,
\end{align}
where $P(0)_i$ is the probability of measuring 0 in the ancilla qubit for the $i$th product module (Eq. \ref{eq:product_layer}),
and $c_i$ are the coefficients in the linear combination of modules.
The trainable parameters are initialized randomly with a standard normal distribution, 
and the training is performed using the Adam optimizer.
As a loss function, we consider the binary cross-entropy with logits (log-loss with a sigmoid activation),
which is well-suited for binary classification tasks. 

For the numerical tests, the training is performed with no mini-batch and a learning rate of $\eta=1$.
We considered 50000 epochs and early stopping with 5000 epochs patience on the validation set F1 score.
We utilized a 10-fold cross-validation method,
and split each training set into $80\%$ training and $20\%$ validation sets.

To evaluate the performance of the architecture,
we consider accuracy as well as the F1 score, the latter being a more robust metric for unbalanced datasets.
The accuracy is defined as the ratio between the number of correctly classified samples over the total number of test samples.
The F1 score is defined as the harmonic mean of precision and recall \cite{pedregosa2018scikitlearn}.

\subsection{Learning real-world data sets} \label{subsec:real_world}

\begin{figure*}[t!]
    \centering
    \includegraphics[width=\linewidth]{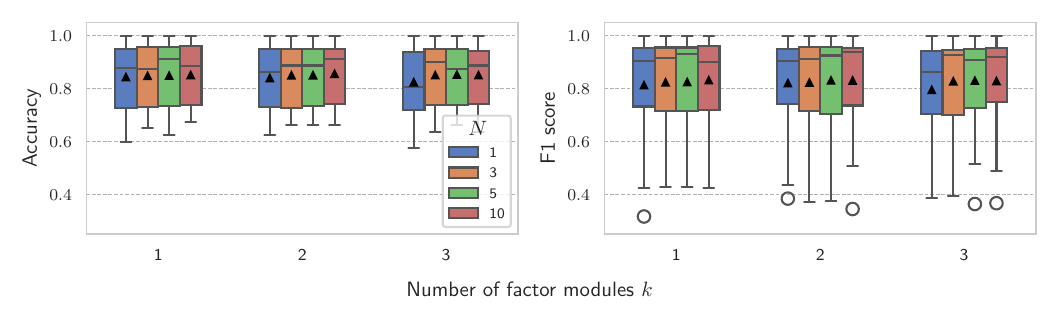}
    \caption{Accuracy and F1 score distribution on the real-world datasets with the PyTorch implementation of the QNN with product layer (see Eq. \ref{eq:product_layer}),
    for increasing number of product modules $N$ and factor modules $k$.
    Each boxplot contains 21 points, each one being the mean value across different folds for each dataset (see Sec. \ref{subsec:training}). 
    Horizontal lines represent the median for all datasets and triangle markers indicate mean values.}
    \label{fig:boxplot_datasets}
\end{figure*}

To test the general capabilities of our architecture (Eq.~\ref{eq:product_layer}), 
we use the QNN as a binary classifier and learn different real-world classical datasets,
originally from the UCI machine learning repository \cite{kelly_home_2024}.
The considered 21 datasets are reported in Table \ref{tab:uci_datasets},
and some of them have been preprocessed to be suitable for binary classification.

\begin{table}[t]
    \small
    \centering
    \begin{tabular}{|c|c|c|}
    \hline
    Dataset & Samples & Features\\
    \hline
    01\_iris\_setosa\_versicolor & 50/50 & 4 \\
    01\_iris\_setosa\_virginica & 50/50 & 4 \\
    01\_iris\_versicolor\_virginica & 50/50 & 4 \\
    03\_vertebral\_column\_2C & 100/210 & 6 \\
    04\_seeds\_1\_2 & 70/70 & 7 \\
    05\_ecoli\_cp\_im & 77/143 & 7 \\
    06\_glasses\_1\_2 & 42/38 & 9 \\
    07\_breast\_tissue\_adi\_fadmasgla & 49/22 & 9 \\
    08\_breast\_cancer & 44/36 & 9 \\
    09\_accent\_recognition\_uk\_us & 63/17 & 12 \\
    10\_leaf\_11\_9 & 14/16 & 14 \\
    11\_banknote\_authentication & 610/762 & 4 \\
    12\_transfusion & 178/570 & 4 \\
    13\_diabetes & 268/500 & 8 \\
    14\_haberman\_survival & 225/81 & 3 \\
    15\_indian\_liver\_patient & 416/167 & 10 \\
    16\_ionosphere & 225/126 & 34 \\
    17\_wdbc & 357/212 & 30 \\
    18\_wine\_quality\_red\_5 & 855/744 & 11 \\
    19\_wine\_quality\_white\_5 & 3258/1640 & 11 \\
    20\_rice\_cammeo\_osmancik & 1630/2180 & 7 \\
    \hline
    \end{tabular}
    \caption{UCI datasets used for the numerical tests in Sec.~\ref{subsec:real_world}.}
    \label{tab:uci_datasets}
\end{table}

We carried out numerical tests with the PyTorch implementation of our proposed QNN architecture,
for different combinations of the number of product modules $N\in\{1,3,5,10\}$ and factor modules $k\in\{1,2,3\}$ in the product layer.
The results are shown in Fig. \ref{fig:boxplot_datasets},
where we report the accuracy and F1 score on the respective test set for all datasets.
We can see that the architecture is able to learn the majority of the considered datasets,
with good values of accuracy and F1 score.

Moreover, we note a slight positive scaling in the prediction performance for the number of product modules $N$,
especially pronounced when passing from a single module to multiple modules.
This is expected, as the expressivity of the network increases with the number of modules, 
and the larger number of modules can represent the same functions as a smaller number of modules.
However, the number of factor modules $k$ in the product layer does not seem to have a noticeable impact on the performance,
as we find similar results for different $k$, suggesting that for these real-world datasets the original architecture (Eq. \ref{eq:standard_qnn}) is sufficient.
The same trend is also observed for individual datasets,
where we find that the performance is similar for different values of $k$. 

The outliers in the F1 score boxplot are related to the transfusion dataset achieving an F1 score around 0.4, 
which is significantly below the scores observed for other datasets. 
However, this dataset has high class imbalance and correlation between two of the four features.

\begin{figure}[t!]
    \centering
    \includegraphics[width=\linewidth]{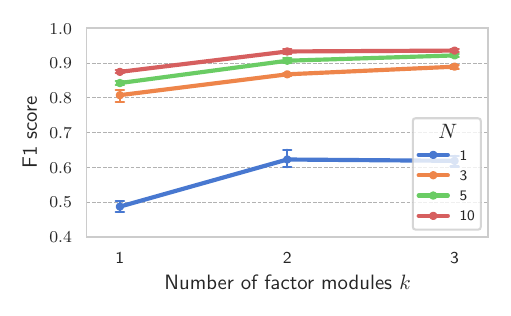}
    \caption{F1 scores for learning the IJCNN1 dataset with our QNN architecture (Eq. \ref{eq:product_layer}). Points represent mean values across different folds, 
    and error bars represent 95\% confidence intervals.}
    \label{fig:ijcnn1}
\end{figure}

For more complex datasets, such as the IJCNN1 dataset \cite{_libsvm_},
we find that the product layer increases the prediction performance.
We merged the original training and test sets into a single dataset 
containing 22 features and 61615 samples,
to utilize the same 10-fold cross-validation procedure as above.
Specifically, in Fig. \ref{fig:ijcnn1} we show the F1 score for the IJCNN1 dataset,
for different combinations of product modules $N$ and factor modules $k$ in the product layer,
obtained with the PyTorch implementation utilizing the same training and testing procedure as above,
with the same parameters.
We considered the F1 score as the only performance metric,
since the dataset is heavily unbalanced (with $\sim 90\%$ of the samples belonging to one of the two classes).
We can observe that the product layer is able to improve the F1 score,
and the performance increases with the number of product modules $N$ and factors $k$ in the product layer,
requiring a number of factor modules $k>1$ to cross the $0.9$ mean F1 score threshold.

\subsection{Learning high-dimensional parity checks} \label{subsec:parity_check}

While the last section showed that a variety of real-world datasets can be easily learned by the standard two-layer feedforward architecture of the SWAP test-based QNN (introduced in Sec.~\ref{subsec:twolayer_nn}), we discuss in this section the parity check function, which is well-known to be hard to learn for many types of neural networks \cite{Dhar2010}. Indeed, we present in Sec.~\ref{subsubsec:analytical_argument} an analytical argument for why the quadratic activation function, used in the SWAP test-based architecture, prohibits the standard two-layer feedforward network to learn, even in principle, the $d$-dimensional parity check function, for $d>2$. In Sec.~\ref{subsubsec:numerics_xor} we then demonstrate with numerical experiments that our proposed modification of the original architecture with product layers (see Sec.~\ref{subsec:productlayer}) allows us to enhance the expressivity of the original network and learn the parity check in arbitrary dimensions. In Sec.~\ref{subsec:spiral} we repeat a similar numerical analysis for another hard-to-learn function: the n-spiral task, for which we find a similar increase in expressivity when using the generalized quantum neural network with product layers.

\subsubsection{Limited expressivity of the QNN without product layer}\label{subsubsec:analytical_argument}

A standard classical two-layer NN with quadratic activation functions does not satisfy the universal approximation theorem \cite{hornik_approximation_1991,pinkus_approximation_1999}. The original quantum neural network architecture (without a product layer), defined in Eq. \ref{eq:standard_qnn}, differs slightly from its classical counterpart, as the bias is encoded in an additional weight parameter (see also discussion in Sec. \ref{subsec:perceptron}). It is thus a priori not clear if the universal approximation theorem equally applies to this QNN for learning classical datasets. However, even if the universal approximation property does not apply, the network can still be suitable for many real-world learning tasks as shown in the previous section.

In this section we give a simple analytical argument for why the standard two-layer feedforward quantum neural network (Eq. \ref{eq:standard_qnn}) will indeed always fail at some learning tasks, even for small input dimensions. Specifically, we show that the $d$-dimensional parity check function (with $d>2$) can never be fully learned due to the quadratic activation function inherent to the SWAP test based architecture. We note that similar results on the limited expressivity of QNN's for the parity check function have been previously discussed in Ref.~\cite{Mingard2024}.

We define the $d$-dimensional parity check function $f_{PC}$ taking as input a $d$-dimensional real vector $\vec{x}$ with non-zero entries and outputs $f_{PC}(\vec{x}) = \mathrm{sgn}\left(\prod_{i=1}^d x_i\right)$, i.e. if $x$ has an even number of negative entries the function output is $+1$, otherwise $-1$. The $d$-dimensional Euclidean space can be separated into the different odd or even orthants. We denote any vector in an even orthant as $\vec{x}^+$, for which  $f_{PC}(\vec{x}^+) =  +1 $ holds. Correspondingly we define $f_{PC}(\vec{x}^-) =  -1 $ for vectors $\vec{x}^-$ in odd orthants. Examples for $\vec{x}^+$ in two dimensions are $(1,1)$ and $(-1,-1)$.

In our analysis we restrict ourselves, without loss of generality, to only the $2^d$  vectors $\{\pm 1\}^d$, one in each orthant. We label all these representative vectors as $\vec{x}^+_i$ and $\vec{x}^-_i$, with  $i=1,...,2^d/2$. The task is then to show that the two-layer feedforward neural network with quadratic activation functions (Eq. \ref{eq:standard_qnn}) is not able to correctly label all $2^d$ of these vectors.

A necessary requirement for the neural network to correctly label all the representative vectors is

\begin{align}
    f(\vec{x}^+_i) > f(\vec{x}^-_j),\quad \forall i,j,
\end{align}

i.e. there exists some threshold value for the output of the neural  network, that correctly distinguishes between the two classes $\vec{x}^+$ and $\vec{x}^-$. The condition $f(\vec{x}^+_i) < f(\vec{x}^-_j)$ is equivalent and can be obtained by simply flipping the signs of all classical coefficients $c_i$ in Eq. \ref{eq:standard_qnn}. The condition above furthermore also implies

\begin{align}
    \sum_i f(\vec{x}^+_i) > \sum_i f(\vec{x}^-_i),
\end{align}

where the sums run over all $2^d/2$ representative vectors in the respective orthants with label $+1$ or $-1$. After inserting the definition of the neural network output $f$ (Eq. \ref{eq:standard_qnn}) and explicitly writing out the bias as the ($d+1$)st entry $w_{j,d+1}$ of the weight vector $\vec{w}_j$ (Eq. \ref{eq:quantum_bias}), we get the following condition:

\begin{align} \label{eq:condition1}
   \sum_i\sum_{j=1}^N  c_j \frac{\left(\vec{x}_i^+ \cdot \vec{w}_j + w_{j,d+1} \right)^2 - \left(\vec{x}_i^- \cdot \vec{w}_j + w_{j,d+1} \right)^2}{||\vec{x}^\prime||^2 ||\vec{w}^\prime_j||^2} > 0,
\end{align}

where we used the fact that $||\vec{x^\prime}|| \equiv \sqrt{||\vec{x}^{\pm}_i||^2 + 1}$ is the same for all $2^d$ representative vectors we consider. Using the following identity for $d\geq3$ (see Appendix \ref{app:proof} for the derivation):

\begin{align}\label{eq:d3_identity}
    \sum_i (\vec{x}_i^\pm \cdot \vec{w}_j)^2 = 2^{d-1}||\vec{w}_j||^2,
\end{align}

we can further simplify our condition to

\begin{align}\label{eq:condition_final}
    \sum_{j=1}^N  \frac{c_j w_{j,d+1}}{||\vec{w}^\prime_j||^2} \vec{w}_j \cdot \sum_i(\vec{x}_i^+ - \vec{x}_i^-) > 0.
\end{align}

We now note that $\sum_i \vec{x}_i^+ = \sum_i \vec{x}_i^- = 0$. This can be easily seen from symmetry or by specifically considering an arbitrary entry in $\vec{x}_i^{\pm}$: If that entry is $+1$, there are $2^{d-2}$ possible configurations of signs in the remaining entries, so that the vector lies in an orthant with label~$\pm$. Analogously there are $2^{d-2}$ other vectors with label~$\pm$ where that specific entry is $-1$. In the sum over all vectors $\vec{x}_i^\pm$ this entry will then cancel to zero. The same argument holds for all other entries in $\vec{x}_i^\pm$.

We thus find that the left side of Eq. \ref{eq:condition_final} evaluates to zero and the inequality can never be fulfilled (in Appendix \ref{app:proof} we show the corresponding condition for $d=2$, which can be fulfilled). From this we        conclude that the set of representative vectors can never be completely distinguished by the two-layer feedforward neural network architecture for $d\geq3$. Note that the above argument equally applies when the representative set is rotated or rescaled arbitrarily. As those representative vectors are a subset of possible inputs in the $d$-dimensional parity check, the function can also not be learned completely in the general case. This argument is independent of the chosen weights, biases and number of modules in the network. We emphasize that this is a fundamental limitation of the architecture due to the quadratic activation function (which also holds for the analogous classical network with biases not encoded in the weights vectors). 

Our numerics in the following section confirm this argument: We find the parity check for $d=2$ to be easily learnable but impossible for $d\geq3$. However, we numerically show that our in Sec.~\ref{subsec:productlayer} proposed generalization of the SWAP test-based quantum neural network can overcome this limitation and learn the parity check function in arbitrary dimensions.

\subsubsection{Numerical results}\label{subsubsec:numerics_xor}

\begin{figure*}[ht!]
    \centering
    \includegraphics[width=\linewidth]{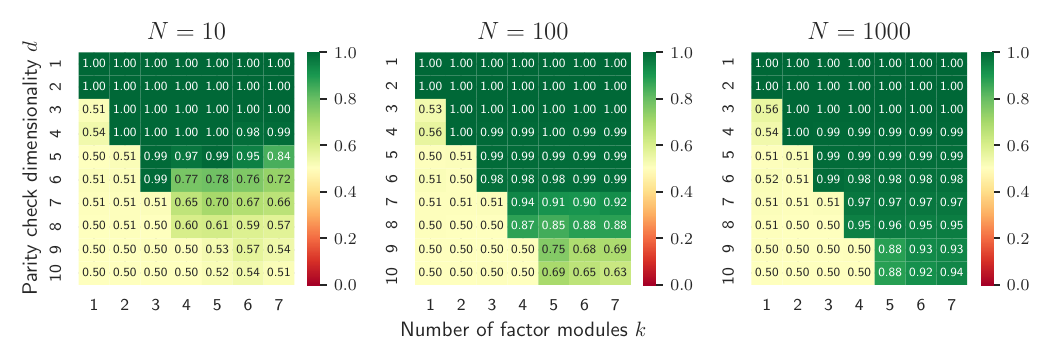}
    \caption{Accuracy results for the parity check data set with the PyTorch implementation of the proposed QNN architecture (Eq.~\eqref{eq:product_layer}). 
    $d$ is the dimension of the parity check data set, 
    $k$ is the number of factor modules in each of the $N$ modules. We generated 1000 samples in each of the $2^d$ respective decision regions.
    Each accuracy point is the maximum achieved accuracy on the test set, 
    obtained by training the network for $50000$ epochs, for learning rates in $\{0.01,0.1,1,10\}$.
    }
    \label{fig:xor_all}
\end{figure*}

We carried out numerical tests on the parity check data set with the PyTorch implementation (see Sec. \ref{subsec:training}), considering input dimensions from 1 to 10.
We considered the generalized architecture define in Eq. \ref{eq:product_layer}, with product modules with repeated inputs for each factor module and a bias encoded in the weights (see Sec. \ref{subsec:perceptron}).
We generated synthetic data sets of the $d$-dimensional parity check function in a balanced manner,
considering $s$ uniformly distributed samples $x \sim U([0,1]^d)$ in each of the $2^d$ different decision regions 
of the $d$-dimensional hypercube, represented by the different combinations of feature signs.
The $d$-dimensional parity check dataset thus contains $s \times 2^d$ samples.
We generated training and test sets independently, with $s$ and $0.2\times s$ samples per region respectively.

From the numerical tests we find that the generalized QNN can classify the parity check dataset at least up to $d=10$
by simultaneously increasing the number $N$ of product modules and the number $k$ of respective factor modules.
We tested for multiple parity check input dimensions $d$,
with different numbers of $k$ factor modules and $N$ product modules. In each case we randomly generated $s=1000$ samples in each decision region. 

Accuracy is considered as the maximum accuracy obtained on the test set,
during $50000$ epochs runs with no early stopping,
considering different learning rates in $\{0.01,0.1,1,10\}$.
We considered a batch size of $256000$ samples (due to GPU memory limit), 
since we observed sensitivity to the gradient calculation.
Hence the training set is processed in a single batch
for the gradient calculation at each epoch, except for the $N=1000$ case for $d=9,10$, 
where we perform $2,4$ optimization steps per epoch respectively.

The achieved accuracies are reported in Fig. \ref{fig:xor_all}. The results demonstrate that the generalized QNN architecture with a product layer can learn the parity check data set for all the considered dimensions to high accuracy,
by increasing the number of factor modules $k$ in the product layer to $k \geq \lceil d/2 \rceil$, corresponding to increasing the degree $2k$ of the effective polynomial activation function.
The number of modules $N$ is also important, as we find accuracy saturation for limited $N$.
Finally, also the number of samples $s$ is important for training, 
as it impacts accuracy at higher dimensions with increased problem difficulty.
Specifically we found accuracy degradation for $s\leq100$ samples per region,
highlighting the sensitivity to gradient calculation for the training and the need for larger training sets.

\subsection{Learning binary n-spiral tasks}\label{subsec:spiral}

In this section we consider the binary $n$-spiral task, which is another classical synthetic dataset that poses a non-trivial learning challenge due to its complex non-linear decision boundaries \cite{Dhar2010}. Similar to the case of the parity check function in the previous section, we find that the standard architecture (Eq. \ref{eq:standard_qnn}) is insufficient for learning higher-order instances of the dataset (all two-dimensional). However, utilizing our proposed architecture with generalized SWAP tests (see Sec. \ref{subsec:productlayer}), we find close to optimal performance when scaling up the network.

\begin{figure*}[htbp]
    \centering
    \includegraphics[width=1\textwidth]{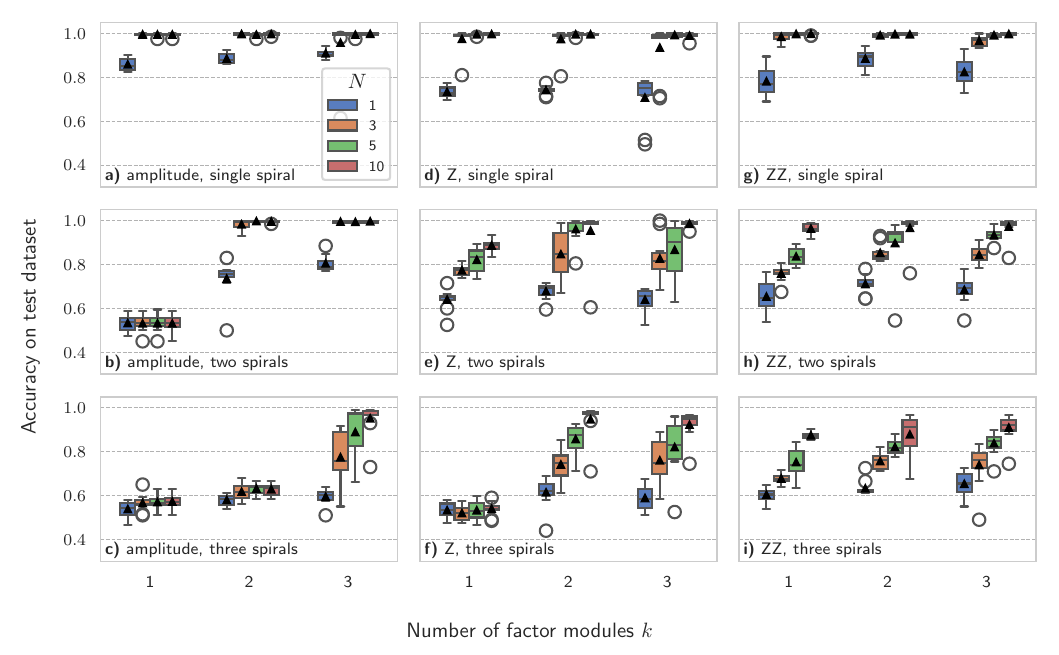}
    \caption{Test accuracy for binary n-spiral classification across three encoding schemes and spiral orders. \textbf{(a-c)} Amplitude encoding (Eq.~\eqref{eq:amplitude_encoding}), \textbf{(d-f)} Z encoding (Eq.~\eqref{eq:U_z}), and \textbf{(g-i)} ZZ encoding (Eq.~\eqref{eq:U_zz}). Rows correspond to increasing spiral order: single (top), double (middle), and triple (bottom) windings around the origin (see Eq.~\ref{eq:spiral_def}). Performance is evaluated as a function of the number of factor modules $k$ (horizontal axis) for different numbers of product modules $N$ (colors). Boxplots represent 10-fold cross-validation results, with horizontal lines showing the median, triangles the mean, and circles indicating outliers.}
    \label{fig:all_test}
\end{figure*}

We generated a collection of binary $n$-spiral datasets with 1000 samples per class for spirals of order one, two and three, where the order denotes the number of times the spirals wind around the origin. In all three cases the generated data is two-dimensional. 
Specifically, we generate the feature vectors $\vec{x}_i^\pm$ for the two classes $+1$ and $-1$ as follows:
\begin{align}\label{eq:spiral_def}
    \vec{x}_i^\pm = \begin{pmatrix}
        \pm r_i \sin(\theta_i) + \epsilon_x\\
        \pm r_i \cos(\theta_i) + \epsilon_y
    \end{pmatrix},
\end{align}
where $r_i = 0.1\times\theta_i$ defines the radius of the spiral, 
and $\theta_i = N_r \times 2\pi \times i / N$ is the angle for sample $i = 1, \ldots, N_s$.
The terms $\epsilon_x,\epsilon_y \sim \mathcal{N}(0,0.04^2)$ represent Gaussian noise.
Here, $N_r$ indicates the spiral order and $N_s = 1000$ denotes the number of samples per class.

The feature vectors $\vec{x}_i^\pm$ are then encoded into a quantum register using amplitude encoding.
In contrast to all of the so far considered datasets, the magnitude of the feature vector encodes crucial information about the two classes for the $n$-spiral task. However, as our architecture is only able to process normalized input vectors (amplitude encoding considers only the angle information but loses the magnitude information due to its underlying quantum circuit, see Eq. \ref{eq:amplitude_encoding}), we have to encode the norm of the two-dimensional feature vectors into a third feature, so that our network effectively trains on a three-dimensional dataset.
Therefore, in a quantum implementation for solving the spiral datasets, the data and weight registers would require two qubits each.

The accuracy results on a test dataset are shown in Fig.~\ref{fig:all_test}a-c, with all hyperparameters chosen as described in Sec. \ref{subsec:training}.
The architecture shows increased performance with a higher number of product modules $N$ and factor modules $k$.
Specifically, increasing the order of the dataset beyond one spiral, requires $k > 1$ to achieve meaningful prediction performance.
Furthermore, increasing $k$ alone is not enough, but has to be combined with an increased number of product modules $N$, similar to what we observed for the parity check function in Sec. \ref{subsec:parity_check}.
For instance, we note that single round spirals (first order) are classified optimally with  $N\geq 2$ and $k=1$ (see Fig.~\ref{fig:all_test}a). However, for second and third order spirals (Figs. \ref{fig:all_test}b and \ref{fig:all_test}c), we require $N\geq 3$, $k=2$ and $N\geq 10$, $k=3$ respectively.

These results, combined with our findings on the parity check function and real-world datasets, indicate that the enhanced expressivity of our generalized SWAP test architecture is not specific to a particular problem structure. Rather, the architecture provides a broadly applicable framework capable of learning diverse classical datasets.

\subsection{Alternative data encoding schemes}\label{subsec:encodings}

Throughout this work, we have employed amplitude encoding to map classical data onto quantum states (Eq.~\eqref{eq:amplitude_encoding}), establishing a direct correspondence between our SWAP test-based QNN and classical feedforward neural networks with polynomial activation functions (see Sec.~\ref{subsec:twolayer_nn}). However, alternative quantum data encoding schemes, such as angle encoding (also known as phase encoding), are widely used in variational quantum algorithms and may offer practical advantages in terms of circuit depth or qubit requirements \cite{Pastorello2023}. It is therefore natural to ask whether the expressivity enhancement achieved through our generalized SWAP test architecture extends to other encoding strategies.

The overall circuit architecture we consider in this section remains identical to that depicted in Fig.~\ref{fig:product_module}: Classical input $\vec{x}$ and weight vectors $\vec{w}_{ij}$ are prepared as quantum states $U(\vec{x})\ket{0}$ and $U(\vec{w}_{ij})\ket{0}$, using unitary encoding circuits $U$, and these states are processed by the generalized SWAP test circuit comprising $k$ factor modules that share a common ancilla qubit. The $N$ measurement probabilities obtained from the product modules are then combined in a purely classical second layer via a weighted sum (analogous to Eq.~\eqref{eq:product_layer}). The only modification with respect to the previous sections lies in the state preparation: instead of amplitude encoding, the unitaries $U$ now implement alternative feature maps.

We consider two representatives: the Z feature map, which encodes data through single-qubit rotations, and the ZZ feature map, which additionally incorporates entangling two-qubit gates. For the Z feature map, we show that the output of a single QNN module reduces to a product of squared cosine functions, reminiscent of classical radial basis function networks that depend on feature-wise differences between inputs and weights. For the ZZ feature map, the entangling gates prevent a simple correspondence with a classical neural network. Despite the loss of the direct analogy to classical two-layer feedforward neural networks, our numerical experiments on the binary n-spiral task demonstrate that the generalized SWAP test circuit can enhance expressivity in certain situations for both alternative encodings, suggesting that the architectural improvements introduced in this work extend beyond the amplitude encoding regime.

\subsubsection{Z encoding}

The Z feature map encodes a real $d$-dimensional classical vector $x$ into a $d$-qubit quantum state
by applying single-qubit rotations.
The unitary encoding circuit $U_Z(\vec{x})$ consists of a layer of Hadamard gates followed by local Z-rotations:
\begin{equation}\label{eq:U_z}
    U_{Z}(\vec{x}) = \left( \bigotimes_{m=1}^d e^{-i Z_m \varphi(x_m) / 2} \right) H^{\otimes d},
\end{equation}
where we set $\varphi(x)=2 x$.
Due to the tensor product structure of this encoding, the SWAP test overlap factorizes into a product of local terms.
The probability of measuring the ancilla qubit in state $\ket{0}$ for product module $i$ with $k$ factor modules is given by (see Appendix \ref{app:z_feature_map} for the derivation):

\begin{align}\label{eq:p0_z}
    \mathbb{P}(0)_i &= \frac{1}{2} \left( 1 + \prod_{j=1}^{k} \left| \braket{0|U_{Z}^\dagger(\vec{x})U_{Z}(\vec{w}_{ij})|0} \right|^2 \right)\nonumber\\
     &= \frac{1}{2} \Bigg( 1 + \prod_{j=1}^{k} \prod_{m=1}^{d} \cos^2 (w_{ij,m} - x_m) \Bigg).
\end{align}

This expression depends only on feature-wise differences $(w_{ij,m} - x_m)$ and factorizes across features, resembling a radial basis function with cosine product activation.

Figure~\ref{fig:all_test}d-f presents our numerical results for the binary n-spiral task using Z encoding. To keep the numerical analysis consistent,
we utilized the same training and testing procedure as for amplitude encoding,
processing the three-dimensional spiral data as described in Sec.~\ref{subsec:spiral}, including the norm as a third feature.
The only modification is in the learning rate, which we set to $\eta=0.1$ in order to achieve better convergence. 

For single-round spirals, the QNN learns successfully with $N>1$ and $k=1$, matching the amplitude encoding results (Fig.~\ref{fig:all_test}a-c). However, for higher-order spirals, the two encodings exhibit notable differences. The Z-encoded QNN achieves reasonable accuracy ($>80\%$), on second-order spirals already at $k=1$, a regime where amplitude encoding fails entirely, though amplitude encoding ultimately achieves higher accuracy for larger $k$ with fewer modules $N$. For third-order spirals, Z encoding reaches near-perfect accuracy with only $k=2$, whereas amplitude encoding requires $k=3$.

\subsubsection{ZZ encoding}
The ZZ feature map encodes a real $d$-dimensional classical vector $x$ into a $d$-qubit quantum state,
by applying both single-qubit rotations and entangling two-qubit gates.
The unitary encoding circuit $U_{ZZ}(\vec{x})$ consists a layer of Hadamard gates followed by local Z-rotations and entangling ZZ-rotations:
\begin{align}\label{eq:U_zz}
    U_{ZZ}(\vec{x}) = \left( \prod_{m=1}^{d} e^{-i Z_m \varphi(x_m) / 2} \right)\nonumber \\
     \left( \prod_{m<n}^{d} e^{-i Z_m Z_n \phi(x_m, x_n) / 2} \right) H^{\otimes d},
\end{align}
where we set $\varphi(x)=2 x$
and $\phi(x, y) = 2 (\pi - x) (\pi - y)$. In contrast to the Z feature map, the entangling gates prevent the SWAP test overlap from factorizing into local terms. Consequently, the ZZ feature map does not have a simple analogy with a classical neural network, and the output depends on correlations between features rather than only on individual feature differences.

Our numerical results for the binary n-spiral task with ZZ encoding are shown in Fig.~\ref{fig:all_test}g-i. Unlike amplitude and Z encoding, increasing the number of factor modules $k$ does not significantly improve performance; instead, accuracy scales primarily with the number of product modules $N$. This suggests that the entangling gates already provide sufficient non-linearity for learning the spiral tasks, reducing the need for the factor modules. Notably, ZZ encoding outperforms both amplitude and Z encoding for second and third-order spirals at $k=1$, though this advantage diminishes for larger $k$ where the other encodings achieve comparable or better results.

Together with the Z encoding results, these findings demonstrate that the generalized SWAP test architecture enhances expressivity across different encoding schemes, not only amplitude encoding. However, the optimal choice of encoding and the degree to which increasing $k$ improves performance depend on the specific problem structure, no single encoding is uniformly superior.

%% file: implementation.tex
To test the performance of our proposed quantum neural network architecture (see Eq. \ref{eq:product_layer}) under noisy conditions, 
we implemented it on real quantum hardware. 
Specifically, we tested the representation capability of the QNN for the three-dimensional parity check function with $N=4$ product modules and $k=2$ factor modules. 
We trained it purely classically, implemented it with the Qiskit library \cite{javadi-abhari_quantum_2024} and then ran the network with the learned weights on the \textit{ibm\_torino} QPU with a \textit{Heron r1} processor. 
For comparison we also performed noiseless simulations of the circuit with the Qiskit Aer simulator.

The transpiled circuit for each of the $N=4$ product modules, including the classical data amplitude encoding and SWAP test, averages a size of $\approx 277$ gates (with $\approx 56$ CZ gates), and an average circuit depth of $\approx 144$.
The transpilation of the SWAP test circuit only, requires on average $\approx 183$ gates (with $\approx 42$ CZ gates) and a circuit depth of $\approx 119$, 
while the amplitude encoding of the input data required on average $\approx 58$ gates (with $4$ CZ gates, one for each 2-qubit data register) and a circuit depth of $9$.
The difference between the sum of data encoding and SWAP test circuit, and the total circuit is due to the additional gates required for qubit routing and topology constraints.
Notably, the single Fredkin gate (CSWAP) requires $41$ gates (with $10$ CZ gates) and a depth of $35$,
which highlights the advantage of a native SWAP test hardware implementation.

We used the same accuracy definition as for the classical surrogate (see Sec. \ref{subsec:training}),
and calculated the accuracy on the test set.
The network achieved an accuracy of 100\% on the test set for the classical surrogate (see Fig. \ref{fig:xor_all}),
and 95\% on the quantum circuit with the noiseless Aer simulator with 8192 shots. 
When running the circuit on the real QPU, we achieved an accuracy of 84\% with 8192 shots.
Despite the slightly worse results on the real QPU due to hardware noise,
the architecture was still able to exhibit high performance and classify the three-dimensional parity check data set with relatively high accuracy.

It should be noted that in the original paper (see Ref. \cite{pastorello_scalable_2024}) on the QNN architecture defined in Eq. \ref{eq:standard_qnn},
the authors proposed a measurement protocol that performs the linear combination of modules
by controlling the number of effective measurements for the $i$-th module to be proportional to $c_i$.
For simplicity we utilized the same number of measurement shots for each module instead, 
and combined each output with the related $c_i$ weight. There is thus potential for further increasing the accuracy when running the QNN on real quantum hardware.

%% file: Conclusion.tex
To summarize, this work presents a comprehensive study of the expressivity of quantum neural networks (QNN) based on the SWAP test quantum circuit across diverse classical datasets. We reviewed the mathematical equivalence between the QNN architecture and classical two-layer feedforward networks with quadratic activation functions under amplitude encoding (established in Ref. \cite{pastorello_scalable_2024}). We then pointed out fundamental limitations stemming from the violation of the universal approximation theorem for polynomial activation functions.

To address these expressivity constraints, we introduced a modified QNN architecture that incorporates generalized SWAP test circuits as building blocks, effectively implementing a classical neural network with a product layer. Crucially, this enhancement preserves the conceptually simple and modular structure of the original QNN, suitable for implementation on current hardware, while significantly expanding the network's representational capacity.

Our extensive evaluation encompassed both real-world datasets (e.g. IRIS and MNIST) and challenging synthetic benchmarks like the parity check function and the binary n-spiral task. The results demonstrate a clear performance dichotomy: while the original QNN architecture successfully represents many real-world datasets, the product layer generalization proves essential for learning the more complex synthetic functions that expose the fundamental expressivity limitations of quadratic activation functions. To rigorously establish these limitations, we provided an analytical argument demonstrating that the original architecture cannot learn the parity check function beyond two dimensions, regardless of network size, a constraint that our generalized architecture overcomes, as evidenced by successful classification of parity check problems in up to ten dimensions.

Beyond amplitude encoding, we also investigated the architecture's performance under alternative quantum data encoding schemes, specifically angle (Z) and ZZ encoding. Our numerical experiments on the binary n-spiral task demonstrate that the generalized SWAP test circuit can increase expressivity under these alternative encodings as well, suggesting that the architectural improvements introduced here are not restricted to a single encoding paradigm.

The practical viability of our approach was validated through implementation on real quantum hardware. A classically pretrained QNN with product layer, deployed on the IBM Torino processor with a Heron r1 chip, achieved $84\%$ classification accuracy on the three-dimensional parity check despite the inherent hardware noise. Our analysis of SWAP test compilation costs highlights the potential advantages of quantum platforms with direct SWAP test implementations, e.g. on photonic or trapped-ion systems, for efficient deployment of these architectures.

Several open questions remain for future investigation. Whether QNNs modeled after classical neural networks offer advantages over standard parametrized quantum circuits remains an active area of research \cite{Wilkinson2022}. Nevertheless, the strong performance of our architecture on classical learning tasks raises compelling questions about its potential for inherently quantum learning applications, such as quantum phase classification or learning properties of quantum states. More broadly, this work establishes a framework for systematically analyzing and enhancing QNN expressivity through correspondence with classical architectures, an approach that could inform the design and evaluation of other quantum machine learning models.

%% file: Acknowledgments.tex
\begin{acknowledgments}
We thank Stefano Azzini and Philipp Hauke for useful discussions.
We acknowledge the use of IBM Quantum services for this work. The authors are solely responsible for the content of this publication. S.N. acknowledges funding by the German Federal Ministry for Education and Research
under the funding reference number 13N16437. 
E.T. was supported by the MUR National Recovery and Resilience Plan (PNRR) M4C1I4.1, 
funded by the European Union under NextGenerationEU.  D.P. was supported by project SERICS (PE00000014) under the MUR National Recovery and Resilience Plan funded by the European Union -- NextGenerationEU. D.P. is a member of the ``Gruppo Nazionale per la Fisica Matematica (GNFM)'' of the ``Istituto Nazionale di Alta Matematica ``Francesco Severi'' (INdAM)''.
Views and opinions expressed are however those of the author(s) only 
and do not necessarily reflect those of the European Union or The European Research Executive Agency. 
Neither the European Union nor the granting authority can be held responsible for them.
This work has benefited from Q@TN, the joint lab
between University of Trento, FBK—Fondazione Bruno
Kessler, INFN—National Institute for Nuclear Physics,
and CNR—National Research Council. We acknowledge
support by Provincia Autonoma di Trento.

S.N and E.T. contributed equally to this manuscript.
\end{acknowledgments}

%% file: appendix.tex
\section{MNIST dataset}\label{app:mnist}

This section evaluates our proposed quantum neural network (QNN) architecture using the MNIST database of handwritten digits \cite{lecun_gradientbased_1998}. MNIST presents a significantly larger input feature space compared to the other real-world datasets discussed in Sec. \ref{subsec:real_world}. Specifically, each sample is a grayscale image comprising $28\times 28 = 784$ pixels.

\begin{figure}[htbp]
    \centering
    \includegraphics[width=\columnwidth]{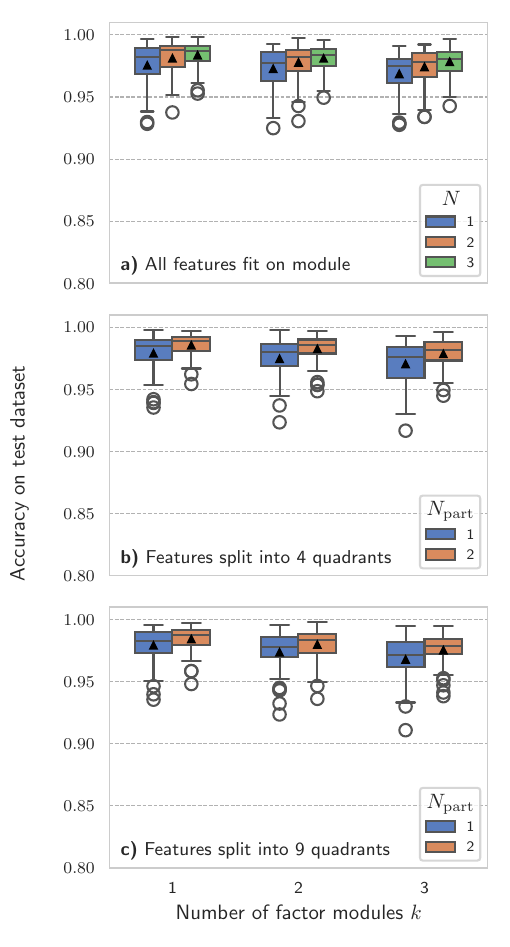}
    \caption{Accuracy results for binary classification after training our generalized QNN (Eq. \ref{eq:product_layer}) on the MNIST dataset. Each boxplot contains $45$ data points corresponding to all possible combinations of digit pairs. In each case we consider a random subset of $10\%$ of the full dataset (hyperparameters discussed in Sec. \ref{subsec:training}; early stopping using accuracy on validaton set). \textbf{a)} Each of the $N$ product modules gets the whole feature vector as input. \textbf{b)} (\textbf{c)}) The feature vector is split into 4 (9) equal image quadrants (see also Sec. \ref{subsec:scalibility}). Each quadrant is the input to $N_{\mathrm{part}}$ product modules, with $k$ factor modules each.} 
    \label{fig:mnist_full_acc}
\end{figure}

We performed binary classification for all 
 unique digit pairs within the dataset. Our initial experiments, depicted in Fig. \ref{fig:mnist_full_acc}a, address the scenario where the complete 784-feature vector can be processed by a single product module (as defined in Eq. \ref{eq:product_layer}). This setup is analogous to the configurations discussed in Sec. \ref{sec:numerics}. The results show that all digit pairs can be classified with high accuracies on a test set exceeding $92\%$. Given the balanced nature of the dataset, the F1 scores are similar. The lowest classification performance was observed for the digit pairs 7-9 and 4-9. Furthermore, a slight improvement in accuracy was noted when increasing the number of product modules $N$. In contrast, increasing the number of factor modules $k$ within each product module did not yield noticeable performance gains for this dataset.

Processing high-dimensional datasets like MNIST on current quantum hardware may necessitate partitioning the input features across multiple modules, a strategy proposed in Ref. \cite{pastorello_scalable_2024} and briefly reviewed in Sec.~\ref{subsec:scalibility}. For the MNIST dataset, amplitude encoding the entire feature vector into a quantum register would require 10 qubits. Such requirements can quickly encounter hardware limitations, for instance, with potential native SWAP test implementations on photonic \cite{Fiorentino2005, Kang2019, Baldazzi2024} or trapped ions \cite{Linke2018, Nguyen2021} quantum computing platforms .

To address this, we compare the results from the full-feature configuration (Fig. \ref{fig:mnist_full_acc}a) with scenarios where the input features are divided into 4 (Fig.\ref{fig:mnist_full_acc}b) or 9 (Fig.\ref{fig:mnist_full_acc}c) equal, non-overlapping spatial partitions. Each partition corresponds to a distinct region of the original image. For example, with 4 partitions, the image is divided into four $14\times14$
sub-images (top-left, top-right, bottom-left, bottom-right). Similarly, 9 partitions correspond to nine $7\times7$
sub-images arranged in a grid.

In these partitioned configurations, each image partition is fed as input to $N_{\mathrm{part}}$ different product modules. Within each such product module, all $k$ factor modules also receive this same input partition. Overall the network is then composed of $\#\mathrm{Partitions} \times N_{\mathrm{part}}$ product modules with $k$ factor modules each.

For both scenarios with partitioned feature vectors we don't observe a noticeable performance decrease in classification accuracy. Like in the case where the full feature vector fits onto a single module, we see a slightly improved performance when increasing the number of modules ($N_{\mathrm{part}}$), while increasing the number of factor modules $k$ has no significant effect.

To summarize, even when quantum hardware might not have enough qubits to encode large feature vectors into single modules, splitting the features onto multiple modules does not necessarily reduce performance when learning classical datasets. It remains an open questions whether similar conclusions hold for learning quantum data.

\section{Derivation of Eq. \ref{eq:d3_identity}}\label{app:proof}

In this Appendix we explicitly calculate
\begin{align}
    \sum_i (\vec{x}_i^\pm \cdot \vec{w}_j)^2 = 2^{d-1} ||\vec{w}_j||^2,
\end{align}

which was used in deriving Eq. \ref{eq:condition_final} in Sec. \ref{subsubsec:analytical_argument} and holds for $d>2$.
We also discuss the case $d=2$ and show the necessary condition (analogous to Eq. \ref{eq:condition_final}) for learning the parity check in 2 dimensions.

To start, we can expand the above sum to 

\begin{align}
    \sum_i (\vec{x}_i^\pm \cdot \vec{w}_j)^2 = &\sum_i\Bigg( \sum_{k=1}^d w_{j,k}^2   \nonumber \\ 
     &+\sum_{1\leq m < n \leq d}2 x^\pm_{i,m} x^\pm_{i,n} w_{j,m} w_{j,n}\Bigg).
\end{align}

Recall that the sum $\sum_i$ goes over the $2^{d-1}$ different representative vectors with label $+1$ or $-1$ respectively. The first part of the expansion can thus be written as $\sum_i ||\vec{w}_j||^2 = 2^{d-1}||\vec{w}_j||^2$. The second part of the expansion can be written as

\begin{align}\label{eq:app_sumexp}
    \sum_{1\leq m < n \leq d}2  w_{j,m} w_{j,n} \sum_i x^\pm_{i,m} x^\pm_{i,n}\nonumber \\
    = \sum_{1\leq m < n \leq d}2  w_{j,m} w_{j,n} S^\pm(d),
\end{align}

where the sum $S^\pm(d)$ does not depend on the indices $m, n$ (due to symmetry) and only on the dimension $d$ of the problem and the label of the representative vectors. For $d > 2$ the sum evaluates to

\begin{align}\label{eq:sum_d3}
    S^\pm(d>2) = (+1)2^{d-2} + (-1)2^{d-2} = 0.
\end{align}

To see this, we can count for how many of the $2^{d-1}$ vectors, over which the sum $\sum_i$ iterates, the product $x^\pm_{i,m} x^\pm_{i,n}$ has a positive or negative sign: There are two possibilities for the product to be $+1$, i.e. $x^\pm_{i,m} = x^\pm_{i,n} = \pm 1$. For either of the two possibilities, the remaining $d-2$ entries have to be even or odd respectively (depending on the label of $\vec{x}^\pm$), i.e. there are $2^{d-3}$ possible vectors. In total we thus have $2\cdot 2^{d-3} = 2^{d-2}$ vectors in the sum $\sum_i$ where  $x^\pm_{i,m} x^\pm_{i,n} = +1$. The analogous argument then holds for $x^\pm_{i,m} x^\pm_{i,n} = -1$ and we get Eq. \ref{eq:sum_d3}. The original sum in Eq. \ref{eq:app_sumexp} thus also evaluates to zero and we obtain the necessary condition Eq. \ref{eq:d3_identity} for the network to learn the parity check in $d>2$ dimensions.

For $d=2$, the situation is different: Since there are only two entries in $\vec{x}_i^\pm$, the product $x^\pm_{i,m} x^\pm_{i,n}$ can only be positive for $\vec{x}_i^+$, or only negative for $\vec{x}_i^-$. In both cases we sum over the two possible vectors with the respective label and obtain 

\begin{align}
    S^\pm(d=2) = \pm 2.
\end{align}

For the original expansion we can thus write

\begin{align}
    \sum_i (\vec{x}_i^\pm \cdot \vec{w}_j)^2 = 2 ||\vec{w}_j||^2 \pm 4   w_{j,1} w_{j,2}. 
\end{align}

Using this expression in Eq. \ref{eq:condition1}, we obtain the following necessary condition, which needs to be fulfilled for the neural network to be able to learn the parity check in two dimensions:

\begin{align}\label{eq:condition_2D}
    \sum_{j=1}^N  \frac{c_j w_{j,1} w_{j,2}}{||\vec{w}^\prime_j||^2}  > 0.
\end{align}

This can be easily satisfied, as also confirmed by our numerical results in Sec. \ref{subsec:parity_check}, e.g. by choosing $N=c_1=w_{1,1} = w_{1,2} = 1$.

\section{Overlap derivation for Z feature map}\label{app:z_feature_map}
In this section, we derive the SWAP test expression for the Z feature map used in Eq. \ref{eq:p0_z}.
Applying the unitary $U_Z(\vec{x})$ to the initial state prepares
\begin{align}
    \ket{\Phi(\vec{x})} &= U_{Z}(\vec{x})\ket{0}^{\otimes d}\nonumber \\
    &= \frac{1}{2^{d/2}} \sum_{z \in \{0,1\}^d} \exp \left( {-\frac{i}{2} \sum_{m=1}^{d} (-1)^{z_m} \varphi(x_m)} \right) \ket{z},
\end{align}
where $z_m$ is the $m$th bit of the bit string $z$ and $\ket{z}$ the corresponding computational basis state.
The overlap between the Z encoded input state $\ket{\Phi(\vec{x})}$ and a weight state $\ket{\Phi(\vec{w})}$ is computed as
\begin{align}
    & \braket{\Phi(\vec{x})|\Phi(\vec{w})} \nonumber \\
    & \; = \frac{1}{2^d} \sum_{z \in \{0,1\}^d} \exp \left(-\frac{i}{2} \sum_{m=1}^{d} (-1)^{z_m} (\varphi(w_m) - \varphi(x_m)) \right).
\end{align}
The summation over the bit strings factorizes into a product of local sums over each qubit:
\begin{align}
    & \braket{\Phi(\vec{x})|\Phi(\vec{w})}= \nonumber \\ 
    & \; = \frac{1}{2^d} \prod_{m=1}^{d} \Bigg( \sum_{b \in \{0,1\}} \exp \big(-\frac{i}{2} (-1)^{b} (\varphi(w_m) - \varphi(x_m)) \big) \Bigg) \nonumber \\
    & \; = \frac{1}{2^d} \prod_{m=1}^{d} \Bigg( 2 \cos \left( \frac{1}{2} (\varphi(w_m) - \varphi(x_m)) \right) \Bigg).
\end{align}
Substituting $\varphi(x)=2 x$, we obtain
\begin{align} 
    \braket{\Phi(\vec{x})|\Phi(\vec{w})} &=  \prod_{m=1}^{d} \cos (w_m - x_m).
\end{align}
Substituting this result into the expression for the probability of measuring the ancilla qubit in the generalized SWAP test in state $\ket{0}$ yields Eq. \ref{eq:p0_z}.